\documentclass[prl,twocolumn,showpacs,groupedaddress,superscriptaddress]{revtex4}
\usepackage{graphicx}
\usepackage{amsmath,amssymb}
\usepackage{xcolor}

\begin{document}

\title{Relaxation Augmented Free Energy Perturbation}

\author{Ying-Chih Chiang}
\affiliation{School of Chemistry, University of Southampton, Highfield, Southampton, SO17 1BJ, United Kingdom}
\author{Livia B. P\'{a}rtay}
\affiliation{Department of Chemistry, University of Warwick, Gibbet Hill, Coventry CV4 7AL, United Kingdom}
\author{Guanglian Li}
\affiliation{Department of Mathematics, Imperial College London, London, United Kingdom}
\author{Christopher Cave-Ayland}
\affiliation{School of Chemistry, University of Southampton, Highfield, Southampton, SO17 1BJ, United Kingdom}
\author{Marley L. Samways}
\affiliation{School of Chemistry, University of Southampton, Highfield, Southampton, SO17 1BJ, United Kingdom}
\author{Frank Otto}
\affiliation{Department of Chemistry, University College London, 20 Gordon Street, London, WC1H 0AJ, United Kingdom}
\author{Jonathan W. Essex}
\affiliation{School of Chemistry, University of Southampton, Highfield, Southampton, SO17 1BJ, United Kingdom}

\date{\today}

\begin{abstract}
Inspired by the recent development on calculating the free energy change 
via a relaxation process [Nat. Phys. \textbf{14}, 842 (2018)],
we investigate the role of heat released in an irreversible
relaxation following a large perturbation.
Utilizing a derivation without microscopic reversibility,
we arrive at a new free energy estimator that employs a volume term 
to account for missing important rare events.
Applications to harmonic oscillators and particle insertion in Lennard-Jones 
fluid agree well with the (numerical) exact solutions.
Our study hence suggests an alternative interpretation to
the insufficient sampling problem in free energy calculations.
\end{abstract}


\maketitle

In recent years, Non-equilibrium physics has become a hot topic that
encompasses many fields, including active matter~\cite{Active_matter}, 
dissipative dynamics that break time-reversal sysmetry~\cite{Dasbiswas18},
entropy production~\cite{Onsager31,Gallavotti04}, etc.
Free energy calculations through non-equilibrium approaches is also one of them. 
Ever since the Jarzynski equality (JE) was first derived~\cite{Jarzynski97PRL},
several ways to derive it have been published, including
derivation via the Crooks theorem~\cite{Crooks00} and via 
the Feynman–Kac theorem~\cite{Hummer01}.
Later on, different free energy estimators have also been proposed:
for instance Adib's clamp-and-release method rooted in 
microscopic reversibility~\cite{Adib06}, 
and the Non-equilibrium Candidate Monte Carlo (NCMC)~\cite{Nilmeier11}
based on a path-wise detailed balance condition.
Recently, it has been shown that the equilibrium free energy can
be extracted from trajectories of the spontaneous thermal 
relaxation processes~\cite{DRoss18}.
Inspired by this study and by the known interrelation between work and heat
in thermodynamics, we investigate the role of heat released in a
relaxation process, initiated by a single large perturbation,
for free energy calculations.

\begin{figure}
\centering 
\includegraphics[width=0.25\textwidth]{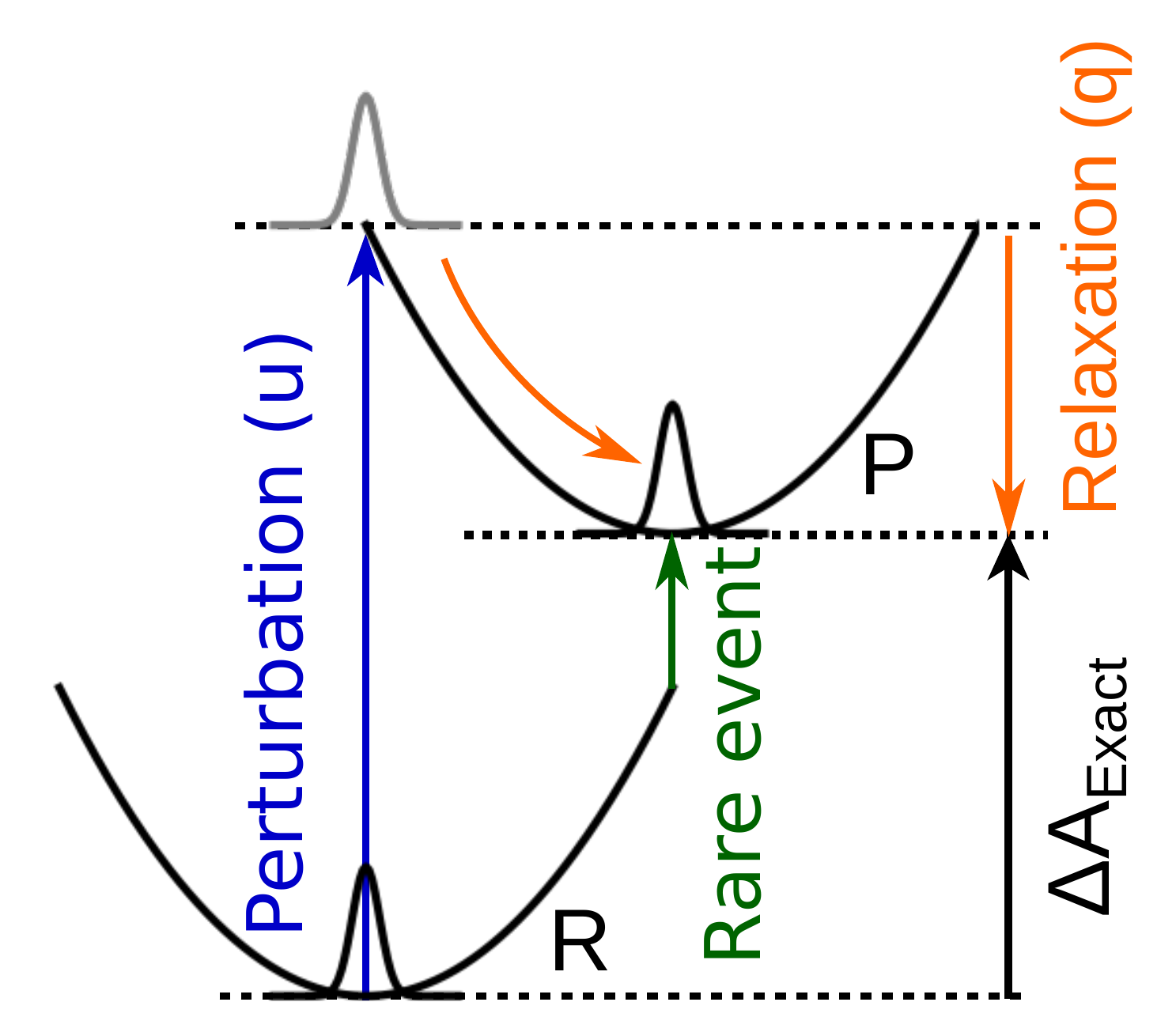} 
\vspace{-0.5cm}
\caption{Schematic potentials of displaced harmonic oscillators. 
The two end states are labeled as $R$ and $P$. The system is initially
at state $R$. Upon turning on the perturbation ($u$) in one single step, 
the system is brought to state $P$ instantaneously.
Afterwards, if the system is allowed to propagate, it will relax spontaneously
by releasing the extra energy as heat ($q$) into the heat bath.
Eventually the system will reach the equilibrium of state $P$.
Traditionally, the rare events, which contribute significantly
to the free energy difference (green arrow), must be sampled to get the
exact free energy difference $\Delta A_{\text{Exact}}$.
By introducing $q$ released from the relaxation process (orange path), 
$\Delta A_{\text{Exact}}$ can also be calculated without sampling
those rare events from state $R$.}
\label{fig:Fig1}
\end{figure}

Consider a system originally at the equilibrium end state $R$.
If the system can be brought into the equilibrium end state $P$
by perturbing its Hamiltonian along a path, then JE states that
the free energy difference between the two end states
can be calculated through the average of the work done to 
the system over all paths~\cite{Jarzynski97PRL}, i.e.
\begin{eqnarray}
\label{eq:JE}
e^{-\beta \Delta A} = \overline{e^{-\beta W}} \,\,\,,
\end{eqnarray} 
where $\Delta A$ is the Helmholtz free energy difference between 
states $P$ and $R$, and $\beta$ is the inverse temperature multiplied 
by the Boltzmann constant ($\beta = 1/k_BT$). 
The work done to the system is denoted by $W$ and
is defined as the accumulated energy change via varying the Hamiltonian along 
the path. The overline indicates the average is performed over all paths,
as $W$ is path-dependent.  Yet, the sampling of possible paths is not trivial: Depending on how the Hamiltonian is perturbed, different
types of paths will be sampled, and different convergence behavior is
thus expected. For instance, if the system is
brought from state $R$ to $P$ instantaneously, i.e. the Hamiltonian 
is perturbed in one single step as in Fig.~\ref{fig:Fig1}, 
the work done to the system is just the potential difference between
the two end states, namely
\begin{eqnarray}
\label{eq:work}
u(x) = U_P(x) - U_R(x) \,\,\,,
\end{eqnarray}
where $u$ denotes the work (perturbation) done to the system,
and $U_P(x)$ and $U_R(x)$ represent the potentials of state $P$
and $R$, respectively.
In this scenario, it was shown that~\cite{Jarzynski97PRL}
the path average is just the ensemble average over $R$,
and the JE is equivalent to the single-step free energy 
perturbation (single-step FEP)~\cite{Zwanzig54}, 
\begin{eqnarray}
\label{eq:sFEP}
\langle e^{-\beta \Delta A} \rangle = \langle e^{-\beta u} \rangle_R \,\,\,,
\end{eqnarray} 
where $\Delta A$ represents the Helmholtz free energy difference between
two end states, and $\langle \cdots \rangle_R$ represents the 
NVT ensemble average over state $R$.
Although Eq.~\ref{eq:JE} and Eq.~\ref{eq:sFEP} are both exact in theory,
in practice they often fail to converge, because the sampling
often misses the important rare events that have ``negative work"
(work smaller than the exact free energy $\Delta A_{\text{Exact}}$), 
see e.g. the green arrow depicted in Fig.~\ref{fig:Fig1}.
To overcome this insufficient sampling problem, calculations 
are usually conducted with a multi-step (slow switch) protocol~\cite{mFEP},
where the system's Hamiltonian is perturbed gradually
through several windows. 
Within each window, the free energy change can again be calculated using 
Eq.~\ref{eq:JE} or Eq.~\ref{eq:sFEP}, and the free energy difference
between state $P$ and state $R$ is given by the accumulated free 
energy change over all windows.
Such operations are termed multi-step FEP~\cite{Chipot10} 
or JE with a slow switch. 

While the rare events (the high energy microstates that have work 
$u \leq \Delta A_{\text{Exact}}$) are not easy to sample, 
exploring the equilibrium microstates of $P$ is straightforward: 
after perturbing the system in one single step (blue line in Fig.~\ref{fig:Fig1}),
propagating the system with a heat bath coupled will result in
a spontaneous relaxation that brings the system to the equilibrium 
of state $P$~\cite{DRoss18}.
Since both rare events and the relaxation explore the equilibrium of $P$, 
we would like to know whether the exact free energy can be calculated
with the relaxation processes instead of the rare events.
During the relaxation process, the extra potential energy is released
as heat ($q$) into the heat bath, so the process violates
the time-reversal symmetry 
(or microscopic reversibility~\cite{Mahan75})~\cite{Dasbiswas18}.
Such a process is by its nature incompatible with 
the Crooks theorem as well as the Adib's nonequilibrium method 
because both of them are rooted in microscopic reversibility.
An alternative derivation outside of those frameworks is thus needed to include
the heat released from the relaxation process.

\begin{figure}
\centering
\hspace{-0.5cm} 
\includegraphics[width=0.24\textwidth]{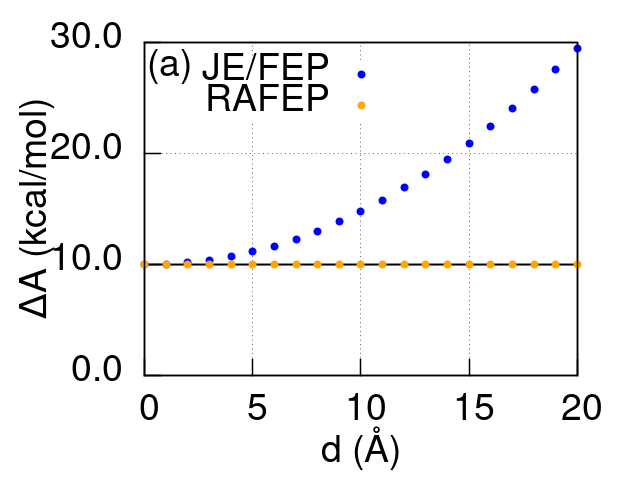} 
\includegraphics[width=0.24\textwidth]{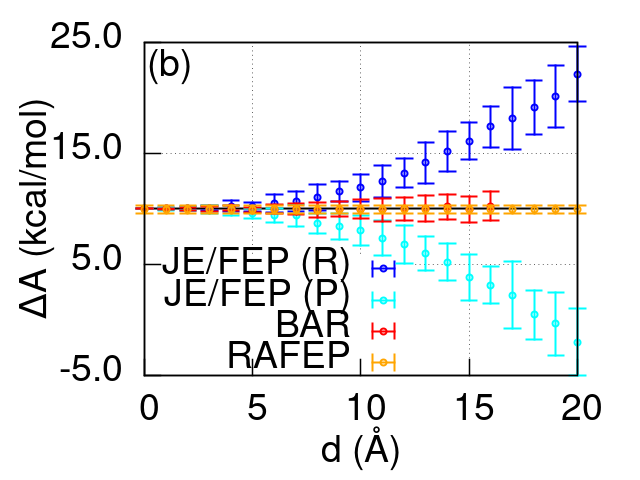} 
\vspace{-0.5cm}
\caption{(a) Free energy difference of displaced harmonic oscillators at 0.1~K.
The force constant $k$ and the minimum potential difference $U_0$
are chosen as 0.1~kcal/mol/\AA{}$^2$ and 10~kcal/mol. 
RAFEP results agree well with the exact solution 
$\Delta A_{\text{Exact}}$ at 10~kcal/mol (black line), 
while the JE/FEP results deviate from $\Delta A_{\text{Exact}}$,
owing to missing important rare events.
(b) Free energy difference of displaced harmonic oscillators at 300~K. 
Potential parameters are the same as in panel (a). 
Again, RAFEP results agree well with $\Delta A_{\text{Exact}}$,
regardless of the value of $d$. In contrast, BAR~\cite{BAR} only works
up to $d=16$~\AA{}. All error bars are calculated by taking the standard deviation
from 50 copies of the calculations, to reflect the size of statistical fluctuation.
}
\label{fig:Fig2}
\end{figure}

We begin with a harmonic oscillator model as depicted in 
Fig.~\ref{fig:Fig1}, at the low temperature limit ($\beta \to \infty$).
For simplicity, the potentials of the two end states $R$ and $P$ are chosen as 
$U_R(x)=\frac{1}{2}kx^2$ and $U_P(x)=\frac{1}{2}k(x-d)^2+U_0$, respectively.  
At $\beta \to \infty$, the probability distribution $P_R(x)$ can be approximated 
by the Dirac delta function as, 
\begin{eqnarray}
\label{eq:Rdistribution}
 \lim_{\beta \to \infty} P_R(x) = \lim_{\beta \to \infty} \frac{1}{\sqrt{2\pi/k\beta}} \exp[ -(\frac{x}{\sqrt{2/k\beta}})^2 ] = \delta(x) \,.
\end{eqnarray}
In the context of sampling, Eq.~\ref{eq:Rdistribution} states that the sampling
is trapped at one single configuration ($x=0$), and
the associated work is given by $u=U_P(0)-U_R(0)=U_0+kd^2/2$.  
Following the conventional definition~\cite{Nilmeier11}, the heat $q$ released in the relaxation process reads,
\begin{eqnarray}
\label{eq:heat}
q(x,x') = U_P(x') - U_P(x) \,\,\,,
\end{eqnarray}
where $x$ and $x'$ are configurations 
before and after the relaxation, respectively.
Similarly, the end point of the relaxation can only be at $x'=d$, 
with the heat $q=U_P(d)-U_P(0)=-kd^2/2$ released.
Since the analytic exact free energy is given by $\Delta A_{\text{Exact}}=U_0$,
it is not difficult to see that $\Delta A_{\text{Exact}}=u+q$, 
leading to a working equation:
\begin{eqnarray}
\label{eq:RAFEP_lowT}
e^{-\beta \Delta A_{\text{Exact}}} = \overline{e^{-\beta (u+q)}} = \langle e^{-\beta (u+q)} \rangle_R  \quad,
\end{eqnarray}
which appears as a natural extension of
Eq.~\ref{eq:JE} and Eq.~\ref{eq:sFEP}, with an extra term $q$ in the average. 
Although this equation only holds at the low temperature limit,
it takes into account the heat released in a relaxation path, 
and hence will be termed ``relaxation augmented free energy perturbation" (RAFEP).
Results of applying Eq.~\ref{eq:RAFEP_lowT} to harmonic oscillators
with different deviation $d$ at 0.1~K are depicted in Fig.~\ref{fig:Fig2}(a).
Compared to JE/FEP with a single step perturbation (via Eq.~\ref{eq:sFEP}),
RAFEP (via Eq.~\ref{eq:RAFEP_lowT}) agrees well with the exact solution 
$\Delta A_{\text{Exact}}$ at 10~kcal/mol, c.f. the blue and orange dots 
to the black line.
This confirms that microscopically the heat released during relaxation 
can contribute to the free energy difference between two end states.
Moreover, calculations with JE/FEP and RAFEP are both performed using
$10^9$ Monte Carlo (MC) steps with a step size 0.1~\AA{}. 
The former spends it all on sampling the 
equilibrium state $R$, but still fails to sample the important rare events. 
The latter first collects 5000 microstates from a simulation of 
$5\times 10^4$ steps, and then relaxes each collected microstates with
a $2\times 10^4$ step-long simulation.
Thus the good performance of RAFEP can only be attributed to introducing 
the relaxation process ($q$) into the free energy calculation.

\begin{figure*}
\centering
\includegraphics[width=0.3\textwidth]{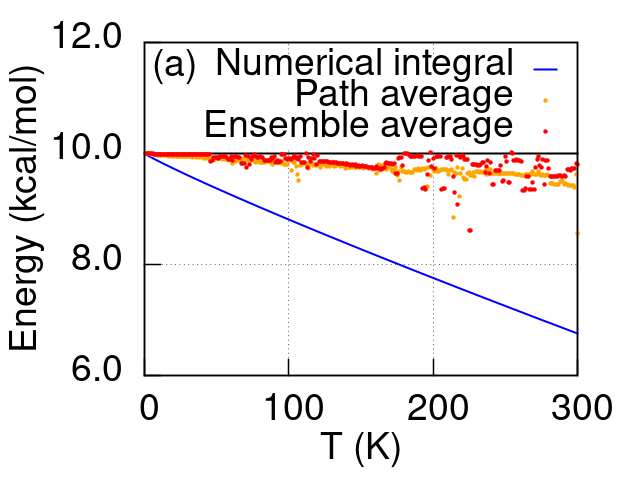}
\includegraphics[width=0.3\textwidth]{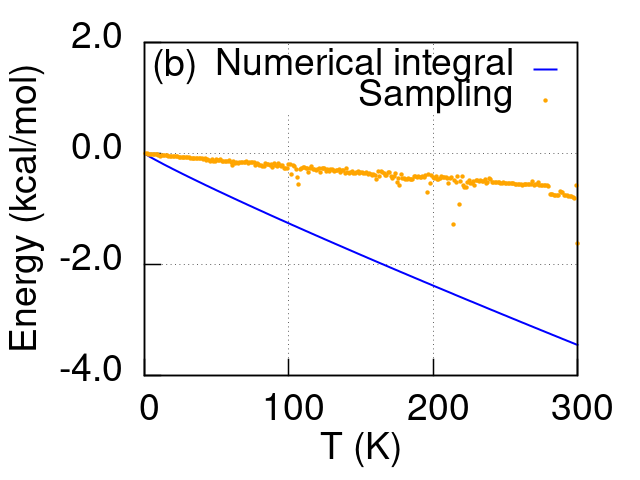}
\includegraphics[width=0.3\textwidth]{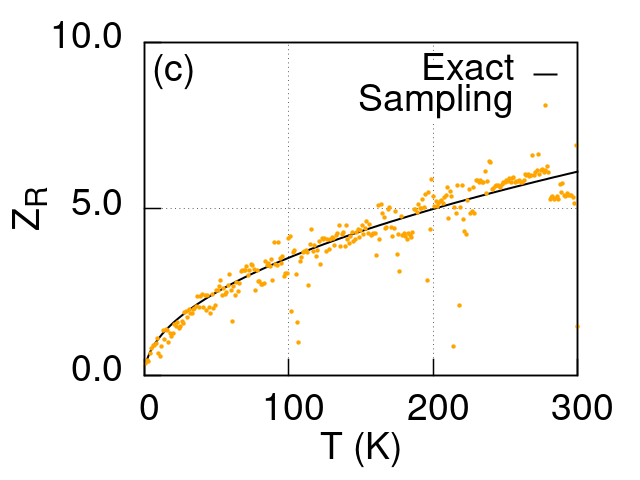}
\vspace{-0.5cm}
\caption{(a) Comparison between the two averages in Eq.~\ref{eq:average}.
Shown are the log value of the average multiplied by $-k_BT$.
The displacement $d$ is fixed at 20~\AA{}. 
All other parameters are the same as in Fig.~\ref{fig:Fig2}(a). 
Both averages deviate slightly from $\Delta A_{\text{Exact}}$ (black line)
as temperature increases.  This deviation becomes more pronounced when
evaluating the average via numerical integration, owing to the choice
of integration boundary (see text).
(b) Evaluation of $-k_BT \langle e^{+\beta U_R} \rangle_R$ via numerical integration and via sampling. Similar deviation as in panel (a) is observed.
(c) Exact partition function of state $R$ ($Z_R$) and the estimated partition
function via sampling of Eq.~\ref{eq:NR2}.  Sampling results in panels (b) and (c)
are taken from the same samples of state $R$ as used for the path average in panel (a).  
}
\label{fig:Fig3}
\end{figure*}

What happens at higher temperature? There the thermal fluctuation
causes the system to populate numerous configurations, leading
to paths with various values of $u$ and $q$.
While the path average remains the same as in Eq.~\ref{eq:RAFEP_lowT}, 
the ensemble average shown in Eq.~\ref{eq:RAFEP_lowT} 
does not yet account for the thermal fluctuation on state $P$,
which would require an extra ensemble average over state $P$.
That is, the two averages are related via
\begin{eqnarray}
\label{eq:average}
\overline{e^{-\beta (u+q)}} = \langle \langle e^{-\beta (u+q)} \rangle_R \rangle_P \quad.
\end{eqnarray}
The double layer ensemble average in Eq.~\ref{eq:average} means that
from every sampled microstate of $R$, multiple relaxation runs
should be performed to sample different microstates of $P$.
Numerical evaluations of the terms in Eq.~\ref{eq:average} are 
depicted in Fig.~\ref{fig:Fig3}(a).
Shown there are the quantities $-k_BT\ln\overline{e^{-\beta(u+q)}}$ and 
$-k_BT\ln\langle \langle e^{-\beta(u+q)} \rangle_R \rangle_P$,
using totally $10^{10}$ relaxation steps. 
The path average distributes these steps into relaxing $5\times10^5$
microstates of $R$ (relaxation length $2\times10^4$ as before),
while the ensemble average repeats relaxing the same 5000 microstates
100 times each. Clearly, when the temperature is higher, the ensemble
average will suffer from a larger fluctuation, cf. the red dots 
shown in Fig.~\ref{fig:Fig3}(a).
Nevertheless, the two averages agree well for $T<160$~K. 
Interestingly, both averages deviate from $\Delta A_{\text{Exact}}$ (black line)
as temperature $T$ increases, and the deviation is even larger if
one evaluates $-k_BT\ln\langle \langle e^{-\beta(u+q)} \rangle_R \rangle_P$ 
by numerical integration (boundary used: -1000~\AA{} to 1000~\AA{}),
as shown by the blue line.
Perhaps this is not so surprising.
Following the definition of work and heat (Eq.~\ref{eq:work} and Eq.~\ref{eq:heat}), the ensemble average can be arranged to
\begin{eqnarray}
\label{eq:average2}
\langle \langle e^{-\beta (u+q)} \rangle_R \rangle_P = \langle e^{-\beta U_P (x')} \rangle_P  \langle e^{+\beta U_R(x)} \rangle_R \quad.
\end{eqnarray}
Evaluation of $\langle e^{-\beta U_P (x')} \rangle_P$ via integration
and via sampling give an identical value, but evaluation of  
$\langle e^{+\beta U_R(x)} \rangle_R$ by the two methods does not,
cf. the orange dots and blue line in Fig.~\ref{fig:Fig3}(b). 
When expressing this term as an integral,
\begin{eqnarray}
\label{eq:NR1}
\langle e^{+\beta U_R} \rangle_R = \int e^{+\beta U_R(x)} P_R(x) dx  \quad,
\end{eqnarray}
where $P_R(x)$ denotes the probability distribution of state $R$,
one sees the problem immediately. 
For an NVT ensemble, $P_R(x)$ is normally taken as the
Boltzmann distribution $e^{-\beta U_R(x)}/Z_R$ with $Z_R$ representing 
the associated partition function.  This results in a diverging 
integral $\int 1 \, dx/Z_R$, if no extra integration boundary is imposed.
As we are using an integration boundary ($-1000$~\AA{} to $1000$~\AA{}),
the blue line depicted in panels (a) and (b) simply reflects this size.
In contrast, the MC sampling always results in a particle position trapped
within $\pm$9~\AA{} from the potential minimum.
This kind of trapped finite sampling is traditionally considered to be the 
problem for evaluating integrals via sampling, e.g. Eq.\ref{eq:sFEP} can 
be derived via integration formalism~\cite{FEP_book} but is problematic
when evaluated via sampling.
Various enhanced sampling methods have been developed to overcome this problem~\cite{Sugita99,Parrinello13}.
However, as the sampling results (orange and red dots) depicted in 
Fig.~\ref{fig:Fig3}(a) follow $\Delta A_{\text{Exact}}$ better than 
the numerical integration (blue line) does, it may be worth formulating 
the problem differently.
Namely, by admitting that the ergodicity is violated in a finite sampling, 
we can describe a finite sampling's behavior with an approximated $P_R(x)$
that reads,
\begin{eqnarray}
\label{eq:PR_approximation}
P_R(x) \approx \frac{e^{-\beta U_R(x)}}{Z_R} \cdot \theta(U_R^*-U_R(x))  \quad,
\end{eqnarray}
where $\theta(U_R^*-U_R(x))$ denotes a Heaviside step function that
caps the population in energy space, based on the maximum energy 
$U_R^*$ encountered during the sampling.
We note that Eq.~\ref{eq:PR_approximation} does not change the actual sampling
but rather the formalism of how a finite sampling relates to 
the associated partition function, i.e. inserting Eq.~\ref{eq:PR_approximation} 
into Eq.~\ref{eq:NR1} results in
\begin{eqnarray}
\label{eq:NR2}
\langle e^{+\beta U_R} \rangle_R &\approx& \frac{\int \theta(U_R^*-U_R(x)) dx}{Z_R} = \frac{V_R}{Z_R}  \quad,
\end{eqnarray}
with $V_R = \int \theta(U_R^*-U_R(x)) dx$ defining the volume of configuration
space that is actually accessed during a finite sampling of $R$.
Similarly for state $P$,
\begin{eqnarray}
\label{eq:NP2}
\langle e^{+\beta U_P} \rangle_P &\approx& \frac{\int \theta(U_P^*-U_P(x)) dx}{Z_P} = \frac{V_P}{Z_P}  \quad,
\end{eqnarray}
where the volume of accessed configuration space
is denoted by $V_P = \int \theta(U_P^*-U_P(x)) dx$,
with $U_P^*$ denoting the maximum energy sampled for state $P$.
Notably, adopting this approximation means that one can estimate
the partition function directly from a finite sampling.
Fig.~\ref{fig:Fig3}(c) shows how $Z_R$ can be sampled by using Eq.~\ref{eq:NR2},
where the volume $V_R$ is here taken from the distance between 
maximum and minimum $x$ of a trajectory.
Following Eq.~\ref{eq:NR2} and Eq.~\ref{eq:NP2}, the RAFEP estimator 
for arbitrary potential and arbitrary temperature reads,
\begin{eqnarray}
\label{eq:RAFEP_1}
e^{-\beta \Delta A_{\text{Exact}}} \approx \frac{\langle e^{+\beta U_R} \rangle_R}{\langle e^{+\beta U_P} \rangle_P} \cdot \frac{V_P}{V_R}  \quad,
\end{eqnarray}
which can be combined with Eq.~\ref{eq:average} and Eq.~\ref{eq:average2}
to give,
\begin{eqnarray}
\label{eq:RAFEP_2}
e^{-\beta \Delta A_{\text{Exact}}} \approx \frac{\overline{e^{-\beta (u+q)}} }{\langle e^{-\beta U_P} \rangle_P \langle e^{+\beta U_P} \rangle_P} \cdot \frac{V_P}{V_R}  \quad.
\end{eqnarray}
Thus, the above two equations are the estimator in two different
forms.  Eq.~\ref{eq:RAFEP_1} combines the samples from two end states for
$\Delta A_{\text{Exact}}$, and is functionally similar to the 
Bennett acceptance ratio (BAR)~\cite{BAR}. 
Eq.~\ref{eq:RAFEP_2} states how the calculation can be performed 
from an equilibrium end state $R$, going through paths including 
an instantaneous perturbation and a subsequent relaxation to arrive at the 
equilibrium of state $P$. 

Results of RAFEP via Eq.~\ref{eq:RAFEP_1} for displaced harmonic oscillators 
at 300~K are depicted in Fig.~\ref{fig:Fig2}(b). 
Calculations are conducted by combining 5000 samples collected over 
$5\times 10^4$ MC steps for state $R$ and for state $P$, 
according to Eq.~\ref{eq:RAFEP_1}.
For comparison, JE/FEP calculations via Eq.~\ref{eq:sFEP} are also
performed using the same amount of sampling, and the associated 
results are further combined through BAR~\cite{BAR}.
Indeed, BAR recovers $\Delta A_{\text{Exact}}$ (10~kcal/mol, black line) 
when $d$ is small, but it still fails when $d \geq 17$~\AA{}.
In contrast, RAFEP via Eq.~\ref{eq:RAFEP_1} outperforms BAR 
in this specific case. 
Although not depicted here, RAFEP calculations via Eq.~\ref{eq:RAFEP_2}
using $10^9$ steps are also performed, and the same results as those 
shown in Fig.~\ref{fig:Fig2}(b) are found.
Either way, the good performance suggests that RAFEP grasps relevant 
physics, even without sampling the rare events.

\begin{figure}
\centering 
\hspace{-0.5cm}
\includegraphics[width=0.2\textwidth]{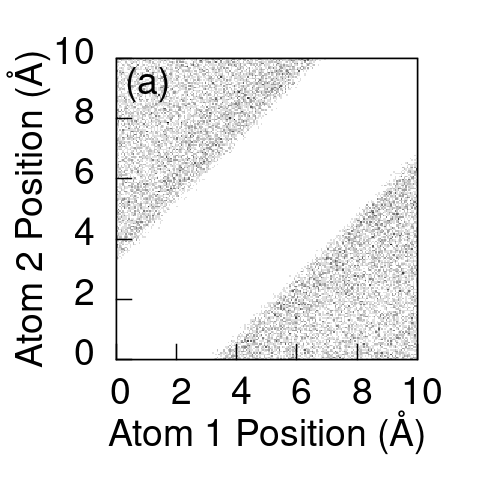} 
\includegraphics[width=0.26\textwidth]{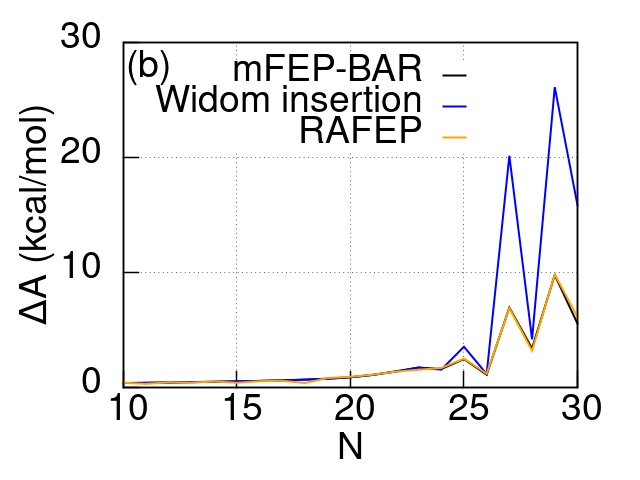} 
\vspace{-0.5cm}
\caption{(a) Trajectory histogram of two Ar atoms in a one-dimensional box.
The diagonal white stripe demonstrates that the particles repel each other
at short interatomic distance, and obviously not all configuration space is
populated within this sampling.  
(b) Free energy change of inserting one Ar atom into $N-1$ Ar atoms inside a 
three-dimensional box. The RAFEP calculation again agrees well with the numerically 
exact reference (mFEP-BAR), while Widom insertion~\cite{Widom63} (see text) 
fails to provide accurate $\Delta A_{\text{Exact}}$ at high particle density.
}
\label{fig:Fig4}
\end{figure}

Next we apply RAFEP to calculate the free energy change upon inserting
one argon (Ar) atom into $N-1$ Ar atoms in a box of size 1000~\AA{}$^3$ 
with periodic boundary conditions. By varying $N$, particle densities 
from gas to solid phase can be explored.
The temperature is set to 85~K, while the Lennard-Jones parameters 
are taken from literature~\cite{Frenkel}: $\epsilon = 0.238$~kcal/mol 
and $\sigma=3.405$~\AA{}. 
The validity of the finite sampling approximation (Eq.~\ref{eq:NR2} and
Eq.~\ref{eq:NP2}) is easily justified by looking at the trajectory histogram 
of two Ar atoms placed in a one-dimensional box, see Fig.~\ref{fig:Fig4}(a)
where the unpopulated white stripe echoes the violation of ergodicity
in a finite sampling ($2\times10^6$ step MC simulation for this example). 
Results of the free energy change upon inserting one Ar atom are shown in 
Fig.~\ref{fig:Fig4}(b). As no analytic solution is available, 
the numerically exact reference is taken from multi-step FEP calculations
combined with BAR (labeled as mFEP-BAR). 
For each $N$, the sampling is performed using totally $2 \times10^9$ MC steps, 
shared between two mFEP calculations (initiated from $R$ and $P$) with 
100 windows each, where $10^7$ steps are employed within each window. 
Note that BAR~\cite{BAR} is mandatory to combine the samples from the two 
mFEP calculations, as their results differ by up to 7.6~$k_BT$. 
When $N \geq 24$, the particle density is comparable with solid Ar,
and inserting one Ar atom would cause a significant reorganization 
of other Ar's positions within the box.  This results in a zigzag shape 
for $\Delta A$, which is also observed in the result of
Widom insertion~\cite{Widom63} (blue curve).
Widom insertion can be viewed as an improved version of Eq.~\ref{eq:sFEP},
where the single uncoupled Ar can be separated from the other $N-1$ atoms 
for brute-force enhanced sampling: here $2 \times 10^7$ steps are used
to sample the $N-1$ atoms, and each sampled configuration is further subjected 
to 100 MC samples of the single Ar. 
In total, the Windom calculation also utilizes $2 \times10^9$ MC steps, but
the method does not reproduce $\Delta A_{\text{Exact}}$ at high particle density,
owing to missing relevant rare events where the $N-1$ Ar atoms
should reorganize to accommodate the inserted atom.
In contrast, RAFEP via Eq.~\ref{eq:RAFEP_1} 
(orange curve) follows closely the numerical reference, even 
when $N>24$. This good performance again confirms that
RAFEP indeed grasps the relevant physics.
Notably, for RAFEP the ensemble averages are calculated with $10^7$ samples
for each end state, which is repeated 10 times to account for statistical
fluctuations, while $V_R$ and $V_P$ are now calculated via
nested sampling~\cite{Skilling04, pymatnest, pymatnest_paper, Livia10},
each with operations equivalent to $0.288 \times 10^9$ MC steps. 
Hence, each RAFEP calculation employs $0.776 \times 10^9$ MC steps 
-- less than half needed for mFEP-BAR.

We have presented a new estimator that is motivated from trying to incorporate 
the heat released during the relaxation process following a single step perturbation, into free energy calculations.
To avoid the limitation of employing microscopic reversibility,
a new derivation based on observing the finite sampling behavior 
is presented. This results in the new estimator RAFEP, which explicitly
admits the violation of ergodicity in a finite sampling and utilizes
a volume term to account for this feature.
RAFEP is then applied to displaced harmonic oscillators 
and particle insertion with Ar atoms.
In both examples, it proves to be a valid estimator and is
free from the need of sampling rare events, demonstrating
that RAFEP provides an alternative and conceptually new view on 
the underlying physics.
Although it is outside the scope of the current paper, scaling up 
RAFEP calculations for biological systems may be of great 
interest~\cite{Cournia17}. 
Crucially the performance of RAFEP depends on how fast the volume terms
can be calculated. For lower dimensions, these can be determined 
effortlessly from a trajectory histogram. This approach soon becomes
impossible as its memory usage growths exponentially as the dimensionality
increases. Yet the hope remains in further advancing the
computational methods, such as nested sampling~\cite{Skilling04, Livia10}, 
population annealing~\cite{Christiansen19}, 
and non-equilibrium importance sampling~\cite{Grant19}.




\bibliography{rafep}

\section*{Acknowledgements}
Y-C.C. thanks Prof. Jarzynski for helpful discussions, 
the Royal Society for funding (NF171278), and the iSolution at 
the University of Southampton for computing time on the Iridis5 cluster.
L.B.P. acknowledges support from the Royal Society through a Dorothy Hodgkin Research Fellowship.

\end{document}